# Pressure-induced superconductivity in LaP$_2$ with graphenelike phosphorus layer

Mingxin Zhang[1], Cuiying Pei[1], Bangshuai Zhu[1], Qi Wang[1,2], Juefei Wu[1*], Yanpeng Qi[1,2,3*]

*1.School of Physical Science and Technology, ShanghaiTech University, Shanghai 201210, China*

*2.ShanghaiTech Laboratory for Topological Physics, ShanghaiTech University, Shanghai 201210, China*

*3.Shanghai Key Laboratory of High-resolution Electron Microscopy, ShanghaiTech University, Shanghai 201210, China*

\* Correspondence should be addressed to Y.P.Q. (qiyp@shanghaitech.edu.cn) or J.F.W (wujf@shanghaitech.edu.cn)

**ABSTRACT**

**Materials with graphene-like layers attract tremendous attention due to their electronic structures and superconducting properties. In this study, we synthesized LaP$_2$ polycrystalline and observed a superconducting transition around 30 GPa. The critical temperature $T_c$ increases monotonically with pressure, which is nearing saturation and reaches 7.8 K at 78 GPa. The synchrotron X-ray diffraction experiments confirm the superconducting transition originates from a structure transition to the *P*6/*mmm* phase under high pressure, suggesting the observation of graphene-like phosphorus layers in transition metal phosphides. By first-principles calculations, we provide more evidence for the stability of the graphene-like phosphorus layers in LaP$_2$. Our findings are helpful for the understanding of the LaP$_2$ phase diagram under high pressure, and could shed light on the research of unique structures in transition metal phosphides under high pressure.**

## 1. INTRODUCTION

The distinctive honeycomb layers of graphene lead to remarkable properties, such as excellent mechanical strength [1, 2], high intrinsic carrier mobility [3], and magic-angle twisted superconductivity [4]. The research highlights in graphene ignite extensive investigations into graphene-like layered materials [5-15]. Notably, high $T_c$ superconductivity has been consistently observed in materials containing graphene-like layers. For instance, the planar honeycomb boron layers in metal diborides ($MB_2$) lay foundations for the superconductivity in $MgB_2$ (0 GPa, $T_c \sim 39$ K) [8], $\alpha$-$MoB_2$ (109 GPa, $T_c \sim 32$ K) [6] and $WB_2$ (91 GPa, $T_c \sim 17$ K) [7, 16]. The buckled hexagonal silicon plane in metal disilicides ($MSi_2$) plays a critical role in the electronic structures of $BaSi_2$ and $CaSi_2$ [9, 10, 17], and the flattening of the buckled silicon plane in $CaSi_2$ induces the superconducting transition of $T_c \sim 14$ K after 12 GPa. Hence, it could be intriguing to explore more graphene-like layered materials for unique superconductors with high $T_c$ values, particularly within the realm of light element compounds.

Phosphorus is one of the most crucial elements in crustal abundance, its abundant valence electrons could form compounds with multiple covalent bonds. Besides, the phosphorus becomes a superconductor under pressure with a $T_c$ value of 6-13 K [18, 19]. More interestingly, some metal phosphides also exhibit superconductivity with modulated P structural units. For illustrations, MnP with edge-sharing octahedrons displays $T_c \sim 1$ K under 8 GPa [20], the noncentrosymmetric superconductor $Mo_3P$ has a $T_c$ value of 5.5 K [21], and $Nb_2P_5$ consists of zigzag-type units with a $T_c$ value of 2.6 K [22]. Thus, the phosphides have potential for designing unique superconductors with distinctive structural units, including 1D chains and 2D layers [23-25]. Recently, the theoretical investigations predict that $LaP_2$ transforms to a *P6/mmm* phase with graphene-like phosphorus layers sandwiching the La atoms after 15.4 GPa, which is a superconductor with a calculated $T_c$ value of 19.9 K at 16 GPa [26]. The electron-phonon coupling (EPC) calculations indicate that the $5d/4f$ electrons in La atoms and the $3p$ electrons in P atoms coupled with the phonon modes from graphene-like phosphorus layers contribute to the superconductivity.

Inspired by the above results, we synthesized $LaP_2$ polycrystalline and conducted comprehensive experimental measurements and theoretical calculations to study the structural and electronic properties of $LaP_2$ under high pressure. We observed the emergence of superconductivity up to 7.8 K after the suppression of semiconductor behaviour. *In situ* synchrotron X-ray diffraction (XRD) experiments revealed a structural phase transition into the graphene-like *P6/mmm* phase under pressure. In addition, we utilized the first-principles calculations to compare the stability of both flat and buckled phosphorus layers based on the graphene-like $LaP_2$ phase under high pressure.

## 2. EXPERIMENTAL DETAILS

Polycrystalline powders were synthesized by solid-state reaction from a mixture of lanthanum (La) powder and phosphorus (P) powder, with an initial rate of 1: 2.1. An excess of phosphorus was employed to compensate for the loss of volatilization. Lanthanum block was mechanically ground into small fragments using a file, while phosphorus blocks were crushed into a fine powder in a mortar. The thoroughly mixed La and P powders were then transferred into an $Al_2O_3$ crucible, which was subsequently sealed within an evacuated quartz tube. The quartz tube was placed into a muffle furnace and subjected to a controlled heating profile: the temperature was gradually increased from room temperature to 400°C within 20 hours, maintained at 400°C for 12 hours, and then further increased to 700°C for an additional 20 hours. The resulting polycrystalline powders exhibited a morphology akin to charcoal and displayed slight sensitivity to oxygen, necessitating storage in an inert glove box.

The phase purity and crystal structure of the synthesized polycrystalline were characterized by powder X-ray diffraction (XRD) using Bruker D8 Advance with Cu $K_\alpha$ radiation. Rietveld refinements of the XRD patterns were executed using the open-source software package GSAS II [27]. The morphology features and elemental composition were confirmed using the scanning electron microscope (SEM) and energy dispersive X-ray spectroscopy (EDS).

Electrical transport measurements under pressure were performed using the Physical Property Measurement System (Dynacool, Quantum Design). High pressure was applied by a commercial diamond anvil cell (DAC) (e.g., easylab, Figure S1[28]) equipped with culets of 200 and 400 μm. The experimental setup employed a BeCu gasket and Pt foils as electrodes (van der Pauw method, Figure S1[28]). Considering that the polycrystalline is isotropic, the sample was put into the cBN hole directly without any other pressure-transmitting medium (PTM). Pressure calibration was achieved via the ruby luminescence method at room temperature [29]. *In situ* high-pressure XRD measurements were carried out at beamline 15U in the Shanghai Synchrotron Radiation Facility (SSRF) utilizing a monochromatic X-ray wavelength of λ=0.6199 Å. Mineral oil was employed as a PTM. The optical band gap was determined using a Cary 5000 spectrophotometer (Agilent Technologies).

First-principle calculations were conducted within the framework of density functional theory (DFT) [30, 31], utilizing the Vienna *ab initio* Simulation Package (VASP) [32]. The exchange-correlation interactions were described by the Perdew-Burke-Ernzerhof (PBE) functional based on the generalized gradient approximation formula (GGA) [33, 34], and the projector augmented wave (PAW) method [35] was employed with valence configuration of $3s^2 3p^3$ for P and $5s^2 5p^6 5d^1 6s^2$ for La. A plane-wave kinetic-energy cutoff of 450 eV was applied, and the Brillouin zone was sampled by the Monkhorst-Pack [36] scheme of $2\pi \times 0.03$ Å$^{-1}$. Convergence criteria were $10^{-6}$ eV for total energy and 0.003 eV/Å for the atomic forces, respectively. Phonon spectrum were calculated using the supercell finite displacement method as implemented in the PHONOPY package [37], with 3×3×3 supercells constructed for all structures.

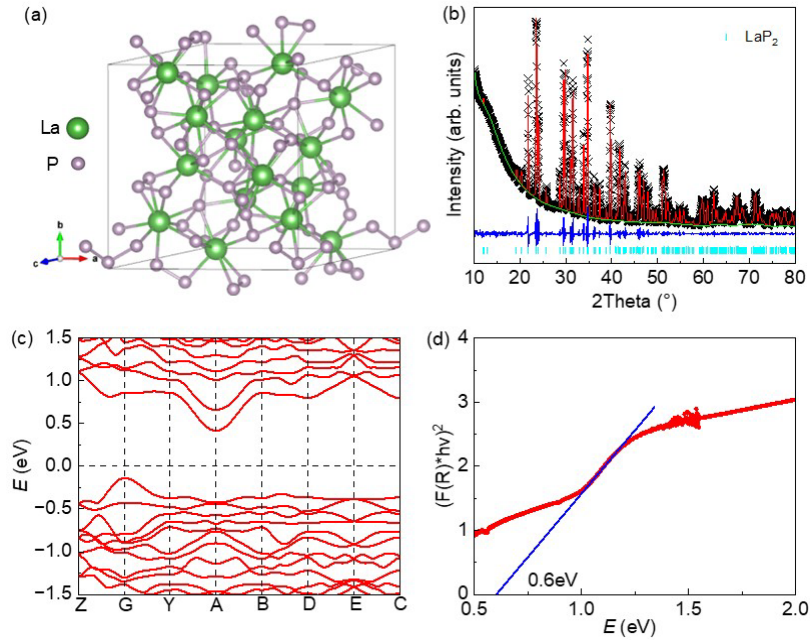

**FIG. 1.** The characteristics of LaP$_2$ polycrystalline. **a.** Crystal structure of LaP$_2$ with *Cc* space group. **b.** The room temperature powder XRD of LaP$_2$ polycrystalline. **c.** Band structure of LaP$_2$ under *Cc* space group. **d.** Determination of indirect band gaps of LaP$_2$ by fitting the optical absorption spectrum.

## 3. RESULTS AND DISCUSSIONS

LaP$_2$ is crystallized in a monoclinic structure with *Cc* space group (No. 9) at ambient pressure [38]. The phosphorus forms bonds around lanthanum atoms, thereby separating the lanthanum atoms from each other (Fig. 1(a)). As demonstrated in Fig. 1(b), the powder XRD pattern of LaP$_2$ polycrystalline shows that all of the peaks can be well refined with *Cc* space group, yielding lattice parameters $a$=12.509 Å, $b$=13.954

Å, $c$=8.889 Å and $\beta$=135.14°, which aligns well with previously reported values [38].

Fig. S2 shows the SEM image and the atomic contents of LaP$_2$ polycrystalline[28]. The EDS results show that the ratio of La and P atoms is 33.93:66.07, in close agreement with the stoichiometry of LaP$_2$. To further understand the properties of LaP$_2$, we calculated the electronic band structures. The first-principles calculation shows that LaP$_2$ was a semiconductor with an indirect bandgap of approximately 0.55 eV (Fig. 1(c)). To experimentally ascertain the band gap size, we conducted optical absorption measurements on the polycrystalline sample, and the measured band gap is of 0.6 eV (Fig. 1(d)), which agrees with our calculations.

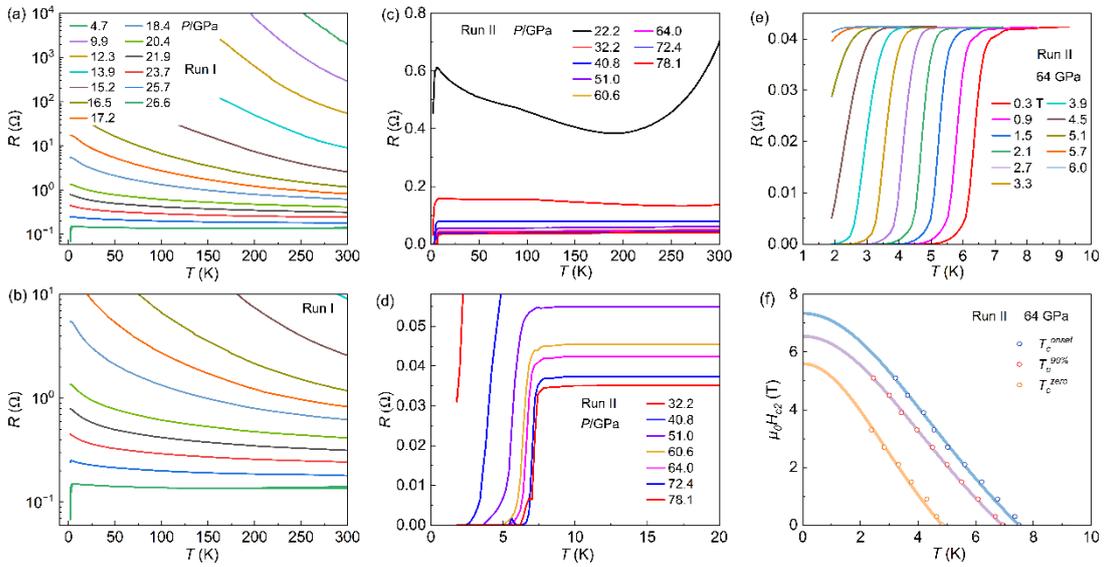

**FIG. 2.** Transport behavior of LaP$_2$ under various pressures. (a)-(b) Temperature dependence resistance of LaP$_2$ up to 26.6 GPa in Run I. (c)-(d) Temperature dependence resistance of LaP$_2$ up to 78.1 GPa in Run II. (e). Temperature dependence resistance of LaP$_2$ under different magnetic fields at 64.0 GPa in Run II. (f) Upper critical field $\mu_0 H_{c2}(T)$ as a function of $T_c^{90\%}$ (90% of the normal-state resistance), $T_c^{onset}$ and $T_c^{zero}$ in Run II.

After the polycrystalline characterization, we performed high-pressure electrical measurements to investigate the electronic properties of LaP$_2$ under high pressure. Fig 2 depicts the temperature-dependent resistance curves under various pressures. Upon cooling, the resistance increases, characterizing a semiconducting behavior (Figs. 2(a) and 2(b)). As the pressure increased, a marked suppression of semiconducting behavior was observed, accompanied by an abrupt resistance reduction at 3.3 K under 26.6 GPa. We employ the thermal activation model to extract the activation energy $E_g$, e.g., $R(T) = R_0\exp(E_g/2k_BT)$, and the fitting results show that $E_g$ decreases rapidly at 10-20 GPa, then tends to be suppressed above 23.7 GPa (Fig. S3[28]). To further corroborate the observed resistance drop at low temperature, we conducted higher pressure measurements up to 78 GPa (Figs. 2(c) and 2(d), Fig. S5[28]). The resistance drop becomes more evident with pressure increasing and reaches zero at 2.5 K under 40.8 GPa, providing experimental evidence of pressure-induced superconductivity. As shown in Fig. 2d, further pressurization will modulate the $T_c$ towards a higher value. When the pressure exceeds 70 GPa, the $T_c$ is nearing saturation and reaches 7.8 K at 78 GPa (Fig.S5 and S6[28]). The different measurements on distinct $LaP_2$ samples (Fig. 2 and S4[28]) yield reproducible results, substantiating the intrinsic superconductivity.

Moreover, we evaluated the robustness of superconductivity under various magnetic fields at 64 GPa (Figs. 2(e) and 2(f)), and the superconductivity was progressively suppressed under external magnetic fields. To determine the upper critical field $\mu_0 H_{c2}(0)$, the empirical Ginzburg-Landau formula $\mu_0 H_{c2}(T) = \mu_0 H_{c2}(T)(1-t^2)/(1+t^2)$ was used to fit the $\mu_0 H_{c2}(T)$ curves, where $t = T/T_c$. We use $T_c^{90\%}$ (90% of the normal-state resistance), $T_c^{onset}$ and $T_c^{zero}$ to calculate the upper critical field. The extrapolated upper critical field $\mu_0 H_{c2}(0)$ of $LaP_2$ reaches 6.5 T ($T_c^{90\%}$) at 64 GPa, and it is below the Pauli limit field $\mu_0 H_P(0)=1.84T_c(\sim 13.2$ T) [39], implying the possibility of s-wave pairing. Based on the relationship $\mu_0 H_{c2} = \Phi_0/(2\pi\xi^2)$, where $\Phi_0 = 2.07 \times 10^{-15}$ Wb, the Ginzburg-Landau coherence length $\xi_{GL}(0)$ is 7.11 nm.

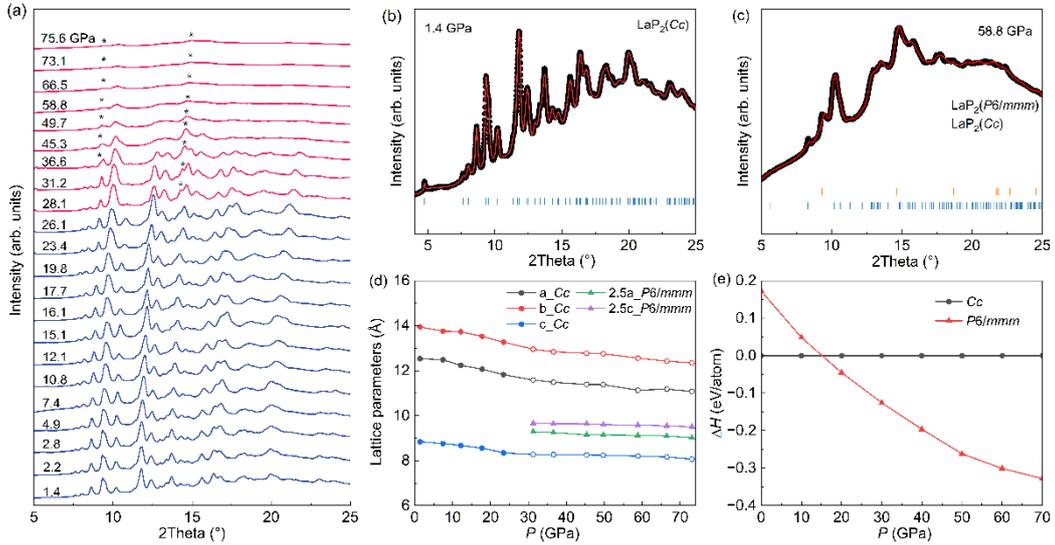

**FIG. 3.** Structure evolution of $LaP_2$ under pressure. **a.** Synchrotron XRD pattern for powder $LaP_2$ sample. The star symbol represents the peaks of phase II (*P6/mmm*), which indicates a partial phase transition under pressure. **b.** Representative fitting of the XRD pattern at 1.4 GPa with $R_{wp}$ = 0.77%. **c.** Representative fitting of the XRD pattern at 58.8 GPa with $R_{wp}$ = 0.75%. **d.** The lattice parameter evolution under pressure of $LaP_2$ with the *Cc* space group and the *P6/mmm* space group. **e.** The enthalpy difference of the *Cc* and *P6/mmm* phase.

The Ref. [26] has conducted theoretical studies on the high-pressure phases as well as the stability of the La-P system, combining first-principles calculations with the crystal structure predictions. The calculation results show that $LaP_2$ undergoes a structure transition from *Cc* phase to *P6/mmm* phase beyond 15.4 GPa, and the predicted *P6/mmm* phase exhibits both thermodynamic and dynamic stability within the pressure range from 15.4 GPa to 66.6 GPa [26]. To uncover the relationship between the crystal structure and the pressure-induced superconductivity in $LaP_2$, we conducted *in situ* high-pressure XRD measurements extending to 75 GPa. As indicated in Fig. 3(a), all diffraction peaks of $LaP_2$ exhibit a continuous shift towards a higher angle. In the low-pressure region, the diffraction peaks can be indexed to the monoclinic *Cc* structure (phase I). A representative fitting of the XRD pattern at 1.4 GPa is shown in Fig. 3(b). Notably, the emergence of two peaks above 28.1 GPa suggests a phase transition (phase II, details are in Fig.S7[28]). With further compression, the intensity of diffraction peaks gradually diminishes. Combining theoretical and experimental results, the synchrotron XRD patterns can be well refined by combining two phases (*Cc* and *P6/mmm*) above 28.1 GPa (Fig. 3(c)). The pressure dependence of lattice parameters was determined by the refined synchrotron XRD patterns. Fig. 3(d) shows the lattice parameters variation as a function of pressure. All these parameters consistently decrease with increasing pressure. *In situ* high-pressure XRD measurements suggest that the emergence of superconductivity is accompanied by the phase transition under pressure, and the electronic structure calculation in Phase II indicates a metallic behavior (Figure S8[28]). Moreover, enthalpy difference analysis (Fig. 3(e)) suggests that the phase with graphene-like phosphorus layers is more energetically favorable under high-pressure conditions, which is in line with Ref. [26].

Additionally, considering the presence of buckled honeycomb sub-lattice in some compounds like $CaSi_2$, $WB_2$ [7, 9, 16, 17], we constructed the $LaP_2$ structure with various buckled phosphorus layers and compared its stability at 50 GPa (Fig. 4(a)). As shown in Fig. 4(b), the height difference 2δ between the upper and lower P atoms reflects the buckling degree of the honeycomb layer, and the relative enthalpy rises with the buckling degree. This suggests the $LaP_2$ structure favours the planar phosphorus layers more than the buckled ones in terms of thermodynamic stability at 50 GPa. To evaluate the dynamic stability, we calculated the phonon dispersion curves of $LaP_2$ with flat (δ=0) and buckled (δ=0.1) phosphorus layers (Fig. 4(c) and 4(d)). The large imaginary frequencies along the whole Brillouin zone in the buckled LaP$_2$ (δ=0.1) indicate its dynamical instability. These results manifest that the phase with graphene-like phosphorus layers in LaP$_2$ is more stable than the buckled honeycomb phase under high pressure.

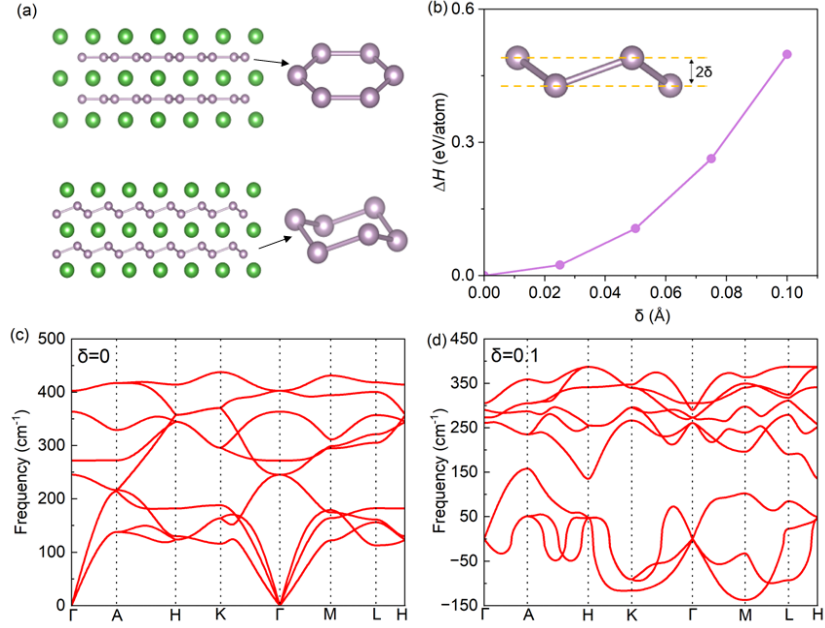

**FIG. 4. a.** The structure of LaP$_2$ under 50 GPa with a flat and buckled phosphorus layer. **b.** The enthalpy difference with various δ, where 2δ represents the height difference within the honeycomb layer. **c.** Phonon dispersion curves of the flat honeycomb phase. **d.** Phonon dispersion curves of the buckled honeycomb phase.

Based on the above results, we summarized the high-pressure phase diagram of LaP$_2$ in Fig. 5(a). LaP$_2$ exhibits a semiconductor behavior in the low-pressure region. The application of pressure induces metallization and the superconductivity transition occurs at around 30 GPa. The emergence of the superconducting state is accompanied by a structural transition from *Cc* to *P6/mmm* phase. The $T_c$ is nearing saturation after 70 GPa with a maximum value of 7.8 K at 78 GPa. Besides, we summarized the $T_c$ values of the reported transition-metal phosphides in Fig. 5(b), and the $T_c$ value of LaP$_2$ is higher than the other reported transition-metal phosphides [5, 20, 22, 25, 40-46]. To further understand the role of graphene-like phosphorus layers, we make comparisons between LaP$_2$ *P6/mmm* and black phosphorus, which consists of a puckered P-atom honeycomb network. In the aspect of crystal structure, the intercalated La atoms separate phosphorus layers, which prevents the bonding between P atoms in the interlayer direction. As for the black phosphorus, the metallic and superconductivity transitions happen under high pressure, while the transitions originate from another simple-cubic (*Pm-3m*) structure under pressure [23, 47]. In terms of superconductivity, $LaP_2$ arises from the comprehensive couplings between La atoms and graphene-like phosphorus layers, while the superconductivity in the pressure induced phase of black phosphorus is more closer to white phosphorus [47]. Besides, black phosphorus crystals can form metal phosphides by intercalating alkali and alkali-earth metals [48], and the $T_c$ near 3.8 K does not depend on the intercalating metal, which indicates that the intercalating metals serve as the electron donor to the phosphorus framework. All these results imply that the graphene-like phosphorus layers could strengthen the EPC between metal atoms and P atoms, which are helpful for the superconductivity in the phosphide materials.

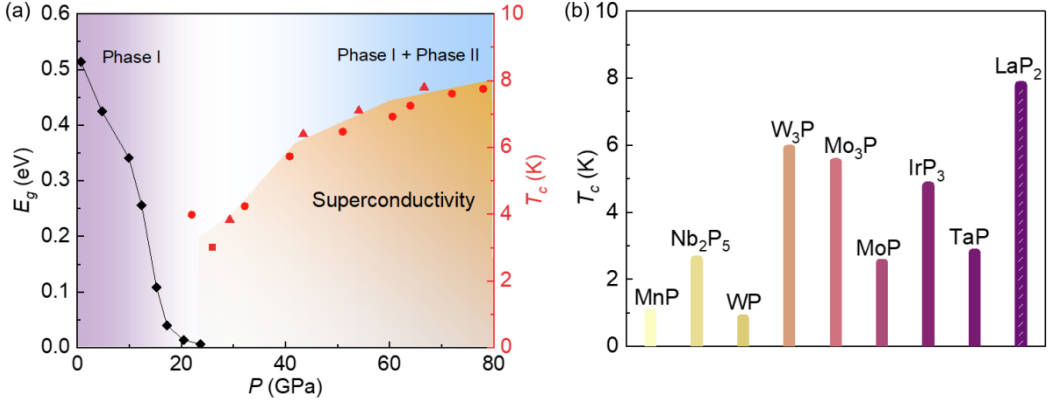

**FIG. 5**. Superconductivity and phase diagram of $LaP_2$ under pressure. **a.** Temperature-pressure phase diagram of $LaP_2$. The colored areas are guides to the eyes, indicating different states. The critical temperature $T_c$ is determined in Runs I-III. **b.** The $T_c$ of $LaP_2$ and typical superconducting transition-metal phosphides.

## 4. CONCLUSION

In summary, we successfully synthesized $LaP_2$ polycrystalline and conducted high-pressure measurements up to 78 GPa. We observed superconductivity after the suppression of the semiconductor behaviors. The superconductivity critical temperature $T_c$ increases monotonously with pressure, attaining a maximum of 7.8 K at 78 GPa. *In situ* high-pressure XRD measurements confirm that the superconductivity transition is associated with a structural phase transition into the *P6/mmm* phase. By constructing buckled phosphorus layers in $LaP_2$, we provide more theoretical evidence for the stability of the *P6/mmm* phase. To our knowledge, this is the experimental report of the graphene-like phosphorus layers in phosphides for the first time. These findings are helpful for understanding the high-pressure properties of LaP$_2$ and will provide insights for studying unique structures in transition-metal phosphides under high pressure.

## ACKNOWLEDGEMENTS

We are grateful to Professor Guochun Yang for stimulating discussion. This work was supported by the National Natural Science Foundation of China (Grant No. 52272265), and the National Key R&D Program of China (Grant No. 2023YFA1607400). The authors thank the support from the Analytical Instrumentation Center (# SPST-AIC10112914), SPST, ShanghaiTech University. The calculations were supported by the HPC platform of ShanghaiTech University. The authors thank the staff from BL15U1 at Shanghai Synchrotron Radiation Facility for assistance during data collection.